\documentclass[a4paper,11pt]{article}
\pdfoutput=1

\usepackage[utf8]{inputenc}
\usepackage[margin=1in]{geometry}
\usepackage[T1]{fontenc}
\usepackage{booktabs,multirow}
\usepackage{bbold}
\usepackage{subcaption}
\usepackage{amsmath}
\usepackage{mathtools,amssymb,braket,dsfont}
\usepackage{url}
\usepackage{graphicx}
\usepackage{multicol}
\usepackage{multirow}
\usepackage{float}
\usepackage[dvipsnames]{xcolor}
\usepackage{slashed}
\usepackage[final,linktoc=page]{hyperref}
\hypersetup{
	colorlinks=true,
	linkcolor=orange,
	citecolor=blue,
	filecolor=magenta,
	urlcolor=blue
}
\usepackage[numbers, sort&compress]{natbib}
\usepackage[utf8]{inputenc}
\usepackage[T1]{fontenc}
\usepackage{comment}
\usepackage{orcidlink}

\begin{document}

\mbox{}\vspace{1cm}

\begin{center}
	 	{\LARGE \bf \boldmath Spontaneous $CP$ Violation and Flavor Changing\\[2mm] Neutral Currents in Minimal SO(10)} \\[1cm]
	{\Large

	Xiyuan Gao\,\orcidlink{0000-0002-1361-4736} $\,^\dag$\footnote{xiyuan.gao@kit.edu}
	}\\[20pt]

 	{
 	$^\dag$ Institute for Theoretical Particle Physics, Karlsruhe Institute of Technology (KIT),\\
        Wolfgang-Gaede-Str. 1, D-76131 Karlsruhe, Germany  \\[5pt]}
\end{center}
\vspace*{1cm}

\begin{abstract}
\noindent
    We explore spontaneous $CP$ violation in the minimal non-super-symmetric $SO(10)$ grand unified theory, with a scalar sector comprising a $CP$-even $45_H$, a $126_H$, and a complex $10_H$. All renormalizable couplings are real due to $CP$ symmetry, and the Kobayashi-Maskawa phase arises solely from complex electroweak vacuum expectation values. The model requires an additional Higgs doublet fine-tuned below 500 GeV and constrains new Yukawa couplings, linking certain flavor-violating (FV) processes. Future proton decay observations may reveal correlated FV decay ratios, offering insights into minimal $SO(10)$.
\end{abstract}

\thispagestyle{empty}

\newpage

\tableofcontents

\section{Introduction}
$SO(10)$ grand unification theory (GUT)~\cite{Fritzsch:1974nn} is one of the most appealing paradigms to understand the Standard Model (SM) of elementary particles and unveil new physics beyond it. Strong and electroweak interactions are successfully unified in renormalizable $SO(10)$ theories without requiring light scalars~\cite{Shafi:1979qb, Bertolini:2009qj}, due to the intermediate scales related to left-right (LR)~\cite{Mohapatra:1974gc, Senjanovic:1975rk} or quark-lepton (QL)~\cite{Pati:1974yy} symmetries. Proton decay is predicted, and the upper limit of proton lifetime is calculable, at least in certain minimal scenarios~\cite{Preda:2022izo, Preda:2024vas}. Furthermore, $SO(10)$ is a complete theory for matter unification. A family of SM fermions plus a right-handed neutrino fit exactly into the spinor representation $16_F$ of $SO(10)$. That explains the seemingly mysterious anomaly cancellation and charge quantization~\cite{Foot:1992ui, Babu:1989tq}. 

Meanwhile, testing $SO(10)$ is highly challenging. Technically, the direct energy scale of all realistic experiments can never reach the unification scale of about $10^{15}$ GeV. An alternative approach is to search the hints at low energy because matter unification implies quarks and leptons behave somewhat similarly. Much effort is focused on numerically fitting the fermion masses and mixing angles, such as~\cite{Dueck:2013gca, Babu:2016bmy, Ohlsson:2019sja, Haba:2023dvo}. However, in our view, these fitted central values should be regarded as predictions because no uncertainties can be identified. For the realistic renormalizable $SO(10)$ GUTs,\footnote{We always take the non-super-symmetric (non-SUSY) framework.} no robust and discriminative results were found till now. Clearly, we need more flavor observables to test $SO(10)$. The difficulty is that $SO(10)$ GUT itself does not require any new sub-TeV particles to mediate flavor transition.

We choose to specify the general $SO(10)$ theory by taking all its renormalizable couplings real. It leads to a more concise theory with fewer free parameters. $CP$ symmetry, the combined transform of charge conjugation $C$ and parity $P$, is enhanced in this limit. $CP$ now serves as a fundamental symmetry of nature~\cite{Landau:1957tp}, neglecting gravitational effects~\cite{Harlow:2018tng}. The $CP$-odd Kobayashi-Maskawa phase~\cite{Kobayashi:1973fv} of weak interactions derives only from the complex vacuum expectation values (VEVs)~\cite{Lee:1973iz}. $CP$ symmetry is spontaneously violated, in the sense that the physical solution does not respect $CP$ symmetry although the Lagrangian does. Spontaneous $CP$ violation (S$CP$V) is not possible in pure SM because one can always rotate the $CP$V phase of the electroweak VEV away, using $U(1)_{\text{Y}}$ gauge redundancy~\cite{Branco:2011iw, Haber:2012np}. In unified theories, the VEVs for $SO(10)$, or intermediate symmetry breaking sometimes get a physical phase, but the low energy theories are almost indistinguishable with the explicit $CP$V one. Such high scale S$CP$V is widely discussed in the literatures~\cite{Chang:1984fx, Chang:1984uy, Patel:2022xxu, Hall:2019qwx, Baldwin:2024bob}.

Interestingly, we notice high scale S$CP$V is impossible in minimal renormalizable $SO(10)$, whose scalar sector only contains a $CP$-even $45_H$, an $126_H$ and a complex $10_H$~\cite{Bajc:2005zf, Bertolini:2012im}. The $126_H$ field indeed contains a large complex VEV, but similar to SM, its phase is also unphysical. S$CP$V is possible and \textit{only} possible together with electroweak symmetry breaking, requiring an additional light Higgs doublet. The new doublet is already contained in $126_H$ or $10_H$, but its mass must be fine-tuned towards the electroweak scale~\cite{PhysRevD.27.1601, Nebot:2018nqn}, just like the SM Higgs doublet. Bearing the fine-tuning, we arrive at a theory without decouple limit. Perturbative unitarity requires the new scalars lie below about 500 GeV~\cite{Nierste:2019fbx}. The low energy theory becomes very similar to what T.~Lee originally proposed~\cite{Lee:1973iz}. The difference is that the Yukawa-type couplings are more constrained than the general two-Higgs doublet models (2HDM)~\cite{Branco:2011iw} because of the LR and QL symmetry of $SO(10)$.

Electroweak scale S$CP$V has been somehow overlooked in the past, partly due to the unavoidable flavor changing neutral currents (FCNC)~\cite{Branco:2006av}. But strictly speaking, the model is on the edge but not ruled out yet~\cite{Nebot:2018nqn, Nierste:2019fbx} because completely eliminating FCNC is not necessary. We treat FCNC as key predictions and advocate greater efforts to measure a set of flavor violating (FV) observables with higher precision, including charged lepton flavor violation (cLFV), neutral meson oscillation, and heavy resonances decaying into FV final states at colliders. These signals do not just shed light on a new Higgs doublet, but also provide more flavor observables related by $SO(10)$. Considering proton decay branching ratios, we have more observables than free parameters. If the FV and proton decay signals are both precisely measured in the near future, we will either get a strong hint on minimal $CP$-conserving $SO(10)$, or directly disprove it.

Our work is structured as follows. In Sec.~\ref{Model}, we revisit the minimal realistic $SO(10)$ theory and discuss its Yukawa sector in the limit of $CP$ symmetry. Then, we intuitively show why the scalar spectrum for electroweak scale S$CP$V contains light 2HDM. Following this, we derive the low energy theory at weak scale and specify its predictions. A more detailed phenomenological analysis is provided in Sec.~\ref{Pheno}. We firstly explain why the theory is not excluded and then show how to test it by comparing the relative strength of a set of experimental observables. Finally, our findings and further discussions are summarized in Sec~\ref{conclu}. In the Appendix, we perform a deeper analysis on the Yukawa sector, illustrate some one-loop GUT-scale corrections, and summarize values for the hadron matrix elements we have used.

\section{Minimal $CP$ invariant $SO(10)$ grand unification theory}
\label{Model}

Three generations of fermionic representation $16_F=(Q_L,u_R,d_R)+(\ell_L,\nu_R,e_R)$ contain all quarks, leptons, and right-handed neutrinos. Then how many new scalar fields must be included in a realistic $SO(10)$ theory? For realistic fermions masses, the simplest choice is a complexified $10_H$ and $126_H$~\cite{Bajc:2005zf}. Neither $10_H$ nor $126_H$ can directly break $SO(10)$ down to the SM gauge group $\text{G}_{\text{SM}}=SU(3)_C\times SU(2)_L \times U(1)_Y$. The minimal path is using an additional $45_H$ to firstly break $SO(10)$ to some intermediate groups, then reduce its rank with $126_H$, and finally get $\text{G}_{\text{SM}}$~\cite{Bertolini:2012im}. Although potentially within the swampland, where the perturbative expansion lies around the edge of breaking down~\cite{Milagre:2024wcg,Jarkovska:2023zwv}, we still choose one $45_H, 126_H$ and a complexified (two real) $10_H=(10_H^1+i 10_H^2)/\sqrt{2}$ as the minimal scalar sector for a renormalizable $SO(10)$ GUT. The symmetry breaking chain is
\begin{equation}
    \begin{aligned}
        SO(10)\times CP ~&~\xrightarrow[M_{\text{GUT}}]{\langle (1,1,1,0) \rangle \in 45_H}SU(3)_C\times SU(2)_L\times SU(2)_R\times U(1)_{B-L}\times CP\\
        ~&~\xrightarrow[M_{R}]{\langle (1,1,3,1) \rangle \in 126_H}SU(3)_C\times SU(2)_L \times U(1)_Y \times CP
        \\
        ~&~\xrightarrow[M_{W}]{\langle (1,2,2,0) \rangle \in 126_H, 10_H} SU(3)_c\times U(1)_{\text{EM}}.
    \end{aligned}
\end{equation}
The $SU(3)_C\times SU(2)_L\times SU(2)_R\times U(1)_{B-L}$ quantum numbers are shown in the parenthesis. Alternatively, the intermediate symmetry can also be $SU(4)_C\times SU(2)_L\times U(1)_R\times CP$, while the other Pati-Salam type breaking patterns are not possible with the minimal scalar sector~\cite{Bertolini:2009qj, Ferrari:2018rey}, unless adding more steps. 
If the intermediate LR (or QL) symmetry breaking scale $M_I$ is around $10^{9}$ GeV (or $10^{11}$ GeV), the gauge coupling unification works perfectly~\cite{Deshpande:1992au,Deshpande:1992em, Bertolini:2009qj}. This result, however, can be relaxed if some of the physical states in $126_H$ and/or $45_H$ are fine-tuned light.\footnote{See~\cite{Preda:2022izo, Preda:2024vas} for a similar theory with $16_H$ and $45_H$.} We therefore do not assume any specific value for $M_I$ but only take it as a high scale, currently not achievable by terrestrial experiments.

If one defines the $CP$ transform for the scalar particles as (with the $SO(10)$ indices implicit),
\begin{equation}
\label{cpdef}
    45_H\rightarrow 45_H, \quad 126_H\rightarrow \overline{126}_H, \quad 10_H\rightarrow 10^{*}_H,
\end{equation}
then all couplings are restricted real. Furthermore, the vacuum of the adjoint representation $45_H$ is real. $126_H$ is a complex representation, but its VEV $\langle (1,1,3,1) \rangle$ can always be chosen to be real by a phase redefinition~\cite{Bertolini:2012im}. As a result, it is impossible to break $CP$ symmetry at high scales with the scalar sector defined in Eq~\ref{cpdef}. In this work, we will not discuss about other alternative definitions of $CP$ symmetry, which may allow pure imaginary couplings. While as a remark, it has been discussed in~\cite{Chang:1984fx} that with an extended scalar of a complex (one $CP$ even, one $CP$ odd) $45_H$, $C$ and $CP$ break together at GUT scale. Whatever, within the minimal $SO(10)$ scenario discussed here, the only possible VEV with a physical phase is from the Higgs $\langle (1,2,2,0) \rangle$. $CP$ is spontaneously violated, only together with electroweak symmetry breaking.

Spontaneous $CP$ violation gives two degenerate vacua, leading to domain wall solutions, a disaster for cosmology~\cite{Zeldovich:1974uw}. A natural way out of the domain wall problem is requiring symmetry nonrestoration in the very early Universe, when the temperature $T\gtrsim M_{W}$~\cite{Weinberg:1974hy, Mohapatra:1979qt, Dvali:1995cc}. Unfortunately, such mechanism is not realistic with only two light Higgs doublets~\cite{Dvali:1996zr}. To keep $CP$ nonrestoration at high temperature, one needs at least a third light Higgs doublet, which requires another fine tuning. A more common solution to the domain wall problem is adding a tiny $CP$-odd perturbation (often called a biased term) to the Lagrangian, so the domain wall is unstable and collapses quickly after its formation~\cite{Zeldovich:1974uw, Vilenkin:1981zs, Gelmini:1988sf}. Such a biased term may derive from effective operators due to quantum gravity, which is expected to violate all global symmetries~\cite{Rai:1992xw, Harlow:2018tng}. The biased term solution is what we choose in this work. In other words, we assume the renormalizable minimal $SO(10)$ theory itself respects $CP$ symmetry exactly, while all the unknown physics beyond grand unification, such as quantum gravity, can break $CP$ but have only negligible effects on the low energy theory, except for destabilizing domain walls. The gravitational waves of domain wall collapse can serve as a smoking gun for electroweak scale discrete symmetry broken, but the amplitudes are, in general, quite small. For potential gravitational wave signals, we refer to~\cite{Hiramatsu:2010yz, Chen:2020soj}.

\subsection{Yukawa sector}
\label{Yukawa}
The $SO(10)$ invariant Yukawa sector reads~\cite{Bertolini:2012im, Mummidi:2021anm, Patel:2022xxu}:
\begin{equation}
\label{YukawaSO10}
    \begin{aligned}
        -\mathcal{L}_Y ~=~ Y_{10} 16_F 10_H 16_F  + \widetilde{Y}_{10} 16_F 10_H^* 16_F  + Y_{126} 16_F \overline{126}_H 16_F  + \text{h.c.}
    \end{aligned}
\end{equation}
$Y_{10}, \widetilde{Y}_{10}$ and $Y_{126}$ are $3\times3$ symmetric matrices according to $SO(10)$ algebra~\cite{Mohapatra:1979nn} and, with the Hermitian condition imposed by $CP$ symmetry, all the elements of $Y_{10}, \widetilde{Y}_{10}$ and $Y_{126}$ are real~\cite{Patel:2022xxu}. As $10_H$ is a complified representation, $10_H$ and $10_H^*$ are independent degrees of freedom, and $\widetilde{Y}_{10}$ is, in general, non-zero. This is different from the scenarios with additional symmetries like $U(1)_{\text{PQ}}$~\cite{Bajc:2005zf}, relaxing the potential constrains from fermions masses and mixing angle's fit~\cite{Dueck:2013gca}. Both $126_H$ and $10_H$ contain two Higgs doublets $\Phi^u$ and $\Phi^d$. So, there are in total four Higgs doublets (though not all light) in the generic basis: $\Phi^i= (\Phi_{10}^d, \widetilde{\Phi}_{10}^u, \Phi_{126}^d, \widetilde{\Phi}_{126}^u)$, where $\widetilde{\Phi}=i\sigma_2\Phi^*$. For simplicity, absorbing all common normalization factors into the Yukawa couplings ($Y_{10}\xrightarrow[]{}\frac{1}{2\sqrt{2}}Y_{10}, \widetilde{Y}_{10}\xrightarrow[]{}\frac{1}{2\sqrt{2}}\widetilde{Y}_{10}, Y_{126}\xrightarrow[]{}\frac{1}{4}\sqrt{\frac{3}{2}}Y_{126}$~\cite{Patel:2022xxu}), the Higgs-fermions interaction reads:
\begin{equation}
\label{YukawaPhi}
    \begin{aligned}
        -\mathcal{L}_Y ~\supset~ 
        & \overline{Q_L}(Y_{10}\Phi_{10}^d-\widetilde{Y}_{10}\widetilde{\Phi}_{10}^{u}+Y_{126}\Phi_{126}^d) 
         d_R 
         +\overline{Q_L}(Y_{10}\Phi_{10}^u+\widetilde{Y}_{10}\widetilde{\Phi}_{10}^{d}+Y_{126}\Phi_{126}^u) 
         u_R \\
        +&\overline{\ell_L} (Y_{10}\Phi_{10}^d-\widetilde{Y}_{10}\widetilde{\Phi}_{10}^{u}-3Y_{126}\Phi_{126}^d)  e_R
        +\overline{\ell_L} (Y_{10}\Phi_{10}^u+\widetilde{Y}_{10}\widetilde{\Phi}_{10}^{d}-3Y_{126}\Phi_{126}^u) \nu_R\\
        +&\frac{1}{2}\overline{\nu_R^c}Y_{\text{126}}\Delta_R^0 \nu_R
        +\frac{1}{2}\overline{\ell_L^c}Y_{\text{126}}\Delta_L \ell_L+ \text{H.c.}
    \end{aligned}
\end{equation}
$\Phi^i$ can develop nonzero complex VEVs: $v_i=(v_{10}^d, v_{10}^{u*}, v_{126}^d,$ $ v_{126}^{u*})$, spontaneously breaking $CP$ together with the electroweak symmetry. The electroweak VEV is defined as $v^2\equiv\sum_{i=1}^4|v_i|^2=(\text{246 GeV})^2$, neglecting small tadpole-induced VEVs of the scalar $SU(2)_L$ triplets ($Y=0,1$) in $45_H, 126_H$.
$\Delta_R^0$ is the neutral component of the $SU(2)_R$ triplet in $126_H$, and its large VEV $\langle \Delta_R^0 \rangle=\langle (1,1,3,1) \rangle$ provides right-handed neutrino masses $M_{\nu_R}$. 
Values of $M_{\nu_R}$ are commonly assumed to be around $10^{13}$ GeV, to give the correct light neutrino masses $M_{\nu_L}$ via the seesaw mechanism. However, $M_{\nu_R}$ can also lie at a lower scale, when the Dirac-type neutrino mass matrix $M_{\nu_D}$ is correspondingly reduced. Smaller values for $M_{\nu_D}$ are allowed since it is not fully aligned with $M_U$. We will explain how to obtain the correct light neutrino mass more explicitly in Appendix~\ref{mixing}.
$\Delta_L$ is the VEV for the $SU(2)_L$ triplet, and also contributes to neutrino masses via type-II seesaw, if sizable. 

At GUT scale, the mass matrices for down-type quarks, up-type quarks, charged leptons, and neutrinos $M_D$, $M_U$, $M_E$, $M_{\nu_L}$ can be derived from Eq~\ref{YukawaPhi}: 
\begin{equation}
\label{fermionsmass}
    \begin{array}{ll}
        M_E=   Y_{10}v_{10}^d-\widetilde{Y}_{10}v_{10}^{u*} -3  Y_{126}v_{126}^d,\quad &
        M_{\nu_R}=   \frac{1}{2}Y_{126}\langle\Delta_R^0\rangle , \\
        M_D=  Y_{10}v_{10}^d -\widetilde{Y}_{10}v_{10}^{u*} +  Y_{126}v_{126}^d,\quad &
        M_{\nu_D} =   Y_{10}v_{10}^u+\widetilde{Y}_{10}v_{10}^{d*} -3  Y_{126}v_{126}^u, \\
        M_U =   Y_{10}^uv_{10}^u+\widetilde{Y}_{10}v_{10}^{d*} +  Y_{126}v_{126}^u, \quad &
        M_{\nu_L} =   -M_{\nu_D}^TM_{\nu_R}^{-1}M_{\nu_D}+\frac{1}{2}Y_{126}\langle\Delta_L\rangle. 
    \end{array}
\end{equation}  
As all Yukawa couplings are real, the nonzero phases of these fermions mass matrices could only come from $v_i$, as the only source of $CP$ violation. When $\widetilde{Y}_{10}=0$, $M_E, M_D, M_U$ are linear dependent and, they thus could be simultaneously diagonalized if neglecting the all quark mixing angles~\cite{Bajc:2002iw}. Numerical fitting shows that this scenario is realistic even considering neutrino oscillation data~\cite{Dueck:2013gca, Ohlsson:2019sja}. However, it is worthy to point out that this is not the intrinsic prediction of $SO(10)$ as it requires additional global symmetries. In general, the three fermions mass matrices (or three Yukawa couplings $Y_{10}, \widetilde{Y}_{10}, Y_{126}$) are not all diagonal and at GUT scale can be parametrized  with
\begin{equation}
    \begin{gathered}
\label{fermionMixing}
    M_D~=~ D^{*}m_D D^{\dag}, \quad
    M_U~=~ U^{*}m_U U^{\dag}, \quad 
    M_E~=~ E^{*}m_E E^{\dag}, \quad
    M_{\nu_L}~=~ N^{*}m_{\nu_L}N^{\dag}, \\
    V_{\text{CKM}}~=~U^{\dag}D, \qquad V_{\text{PMNS}}~=~E^{\dag}N, \qquad V_{E} ~=~ E^{\dag}D .    
\end{gathered}
\end{equation}
$m_D, m_U, m_E, m_{\nu_L}$ are diagonal matrices containing the physical masses for quarks and leptons. $V_{\text{CKM}}$ is the Cabibbo-Kobayashi-Maskawa (CKM) quark mixing matrix and $V_{\text{PMNS}}$ is the Pontecorvo-Maki-Nakagawa-Sakata (PMNS) neutrino mixing matrix. Thanks to $SO(10)$ algebra, the symmetric $Y_{10}, \widetilde{Y}_{10}, Y_{126}$ requires that the mixing matrices for left-handed or right-handed fermions are the same at GUT scale, so the unknown physical matrix is merely $V_E$. In comparison, the general Yukawa potential with four Higgs doublets contains 16 independent general $3\times3$ coupling matrices, instead of just three real symmetric ones in Eq~\ref{YukawaPhi}. This tells us that the flavor structure for fermion-scalar interactions of minimal realistic $SO(10)$ are strongly constrained at tree level, accounting for an important prediction of matter unification itself.

Does Eq~\ref{fermionsmass} predict anything more? One may expect so because there are only 21 physical free parameters (taking one of the VEVs real, and one of the Yukawas diagonal) determining all SM flavor observables. Numerical analysis was performed in exactly the same scenario~\cite{Patel:2022xxu}, and the best-fit solutions were found. However, those should not be interpreted as predictions, because no uncertainties can be quantified. Assuming one finds a million data points consistent with the SM fermions masses and mixing angles, it still only represents a negligible part of the 21-dimension parameter space ($10^{-15}$, conservatively assuming ten points can sample each dimension). Clearly, some (semi)analytical analysis is required, and we have explored this approach. Unfortunately, we find no robust constrains at the end, and some details are shown in Appendix~\ref{mixing}. The main difficulty is that $M_{\nu_L}$ depends on the mass matrices \textit{nonlinearly}, as a result of the seesaw mechanism. With some approximations, we find the mixing angles of $V_E$ can range from zero to large values similar to the neutrino mixing angles. In addition, the VEVs also appear poorly constrained. Although we cannot rule out some potential undiscovered relationships, Eq~\ref{fermionsmass} alone does not seem to yield robust predictions. $SO(10)$ may leave clues about fermions masses and mixing angles, but they are deeply hidden.

So far, we have not included the effects of renormalization group (RG) running, so all results above are only valid at the GUT scale (tree-level). The effects from QCD and possibly scalar self-interactions are expected to be large, but they are fortunately flavor blinded. These corrections can be absorbed into the overall factors, so do not change the physical predictions. The Yukawa couplings are, in general, flavor non-universal, leading to corrections to some GUT-scale predictions at low energy. However, the RG effects are suppressed by the loop factor $1/(4\pi)^2$ so remain at next-to-leading order in a perturbative theory. For simplicity, all our subsequent analysis will be restricted to GUT scale (tree-level) predictions. Some quantitative estimations of the low energy deviations are provided in Appendix~\ref{rge}.

\subsection{Scalar spectrum}
\label{Scalar}
\begin{figure}[t]
  \centering
  \includegraphics[width=0.6\textwidth]{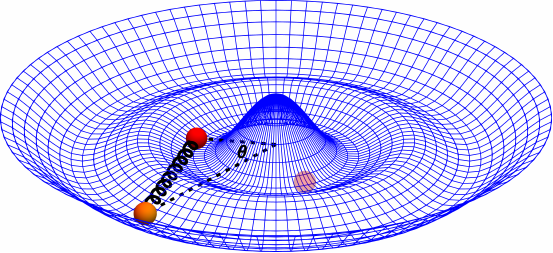}
    \caption{Geometrical illustration of spontaneous $CP$ violation together with $U(1)_Y$, based on the planar diagram of T.Lee~\cite{Lee:1973iz} (FIG-1). The ``Mexican Hat''-like potential itself has rotation ($U(1)_Y$) and reflection ($CP$) symmetries, but the stable physical solution does not, when the balls fall into the valleys and the spring is relaxed. The light-red ball implies there is a mirror solution $\theta\rightarrow-\theta$, and nature has to choose one as the true vacuum. The $\theta=0$ state is invariant under reflection, but not stable. It is a local maximum because the string is contracted and the total energy of the system increases.}
   \label{SCPVgeoView}
\end{figure}
The natural mass scale of all nonchiral particles in minimal $SO(10)$ is $M_{\text{GUT}}$, due to the large VEV of $45_H$. Clearly, the naturalness criterion fails here, because the SM Higgs boson mass is 125 GeV, much lighter than $M_{\text{GUT}}$. It has been realized for a long time, that electroweak scale S$CP$V requires two light Higgs doublets and double fine-tuning~\cite{PhysRevD.27.1601}. This is a model-independent result, and we gave a geometric illustration in Figure~\ref{SCPVgeoView}, based on the planar diagram of T.~Lee~\cite{Lee:1973iz} (FIG-1). The Mexican Hat-like potential has rotation ($U(1)_Y$) and reflection ($CP$) symmetries, but they are broken when the spring-connected two balls fall into the valleys. As shown with the orange and dark red balls, the true vacuum of the system corresponds to the state when the balls are at the bottom of the valleys and the spring is fully relaxed. A mirror vacuum solution is illustrated with the light red ball, corresponding to $\theta\rightarrow-\theta$. These two degenerate solutions are separated by a local maximum, the $\theta=0$ state. This is because the spring is contracted when $\theta=0$, increasing energy of the whole system. As a result, although the $\theta=0$ state preserves reflecting ($CP$) symmetry, it is not the stable solution of the system. Nature has to choose the true vacuum between the $\pm\theta$ states, spontaneously breaking $CP$ symmetry. The rotation redundancy of the whole system corresponds to the Goldstone mode of $U(1)_Y$ broken $G^0$. The massive degrees of freedom can be identified as two radical modes of oscillation in the valleys $h$ and $H$, and one axial mode vibrating along the spring $A$. The values of the masses are the corresponding oscillation frequencies. Within a perturbative theory, that valley curvature and the spring elastic coefficient are both bounded from above by roughly the squared distance between the two peaks (in natural units). Therefore, we have a perturbative unitarity bound in analog to the single Higgs boson case~\cite{Lee:1977yc}: 
\begin{equation}
\label{illuUpperbound}
    \sum_{h,H,A} \omega_\text{osci}^2~=~m_h^2+m_H^2+m_A^2 ~\lesssim~ M_{\text{LQT}}^2 ~=~ (700\sim 800~\text{GeV})^2. 
\end{equation}
Although merely a geometrical illustration, Eq~\ref{illuUpperbound} aligns quite well with the explicit next-to-leading order analysis. If identifying $h$ as the discovered 125 GeV SM Higgs boson, then it is shown in \cite{Nierste:2019fbx} that $m_H$ and $m_A$ are individually bounded by 485 GeV and 545 GeV, so S$CP$V with $U(1)_Y$ indeed has no decouple limit. This is clearly a consistent requirement for the low energy spectrum, so adding more heavy Higgs doublets does not change the scenario, as recently proven in~\cite{Miro:2024zka}. Generalization to the electroweak gauge symmetry $SU(2)_{L}\times U(1)_Y$ is straight forward.

Therefore, the low-energy particle spectrum for minimal $CP$ conserving $SO(10)$ is the same as light 2HDM. But remember, we have in total four Higgs doublets in the generic basis, two in $10_H$ and two in $126_H$. To go to the low energy effective theory, we firstly define the following light states:
\begin{equation}
    h^{\text{SM}}~=~\frac{v_i^*}{v}\rho_i, \quad \widetilde{H}~=~\frac{u_i^*} {u}\rho_i,\quad G^0~=~\frac{v_i^*}{v}\eta_i, \quad\widetilde{A}~=~\frac{u_i^*}{u}\eta_i, \qquad i=1,2,3,4, 
\end{equation}
where 
\begin{equation}
    \Phi_i= \left(\begin{array}{c}
         \phi_i^+  \\
         (v_i+\rho_i+i \eta_i)/\sqrt{2} 
    \end{array}\right), \quad u_i=v_i^*-\left(\sum_{j=1}^4 v_j^{*2}\right)\frac{v_i}{v}, \quad u=\sum_{i=1}^4 |u_i|^2.
\end{equation}
So $h^{\text{SM}}$ and $\widetilde{H}$ live in the plane spanned by $v_i$ and $v_i^*$, the two degenerate vacuum solutions. It's straight forward to check $\sum_i^4 v_i^* u_i=0$, so $h^{\text{SM}}$ is orthogonal to $\widetilde{H}$ (and similarly for $G^0$ and $\widetilde{A}$). In addition, there are two other linear-independent combinations of $\rho_i$ (and $\eta_i$), accounting for the superheavy neutral states irrelevant to phenomenology. Thus, $\rho_i, \eta_i$ can be represented by
\begin{equation}
\label{mixHeavy}
\begin{aligned}
    \rho_i~&=~\frac{v_i}{v}h^{\text{SM}}+\frac{u_i}{u}\widetilde{H}+\text{superheavy States},\\
    \eta_i~&=~\frac{v_i}{v}G^0+\frac{u_i}{u}\widetilde{A}+\text{superheavy States}. 
\end{aligned}
\end{equation}

What follows next is the same as the general 2HDMs~\cite{Branco:2011iw, Inoue:2014nva}. $G^0$ can be identified as the massless Goldstone mode, while the other two $CP$-even components $h^{\text{SM}}, \widetilde{H}$ and one $CP$-odd component $\widetilde{A}$ are still not the mass eigenstates. In general, they mix with each other, and the physical states can be parameterized with
\begin{equation}
\label{HiggsMix}
    \left(\begin{array}{c}
         h  \\
         H  \\
         A  \\
    \end{array}\right)=
    \left(\begin{array}{ccc}
        1 & 0 &0 \\
        0 & c_{\alpha_c} & s_{\alpha_c}\\
        0 & -s_{\alpha_c} & c_{\alpha_c} \\
    \end{array}\right)
    \left(\begin{array}{ccc}
        c_{\alpha_A} & 0 & s_{\alpha_A} \\
        0 & 1 & 0\\
        -s_{\alpha_A}  & 0 & c_{\alpha_A} \\
    \end{array}\right)
    \left(\begin{array}{ccc}
        c_{\alpha_H} & s_{\alpha_H}  & 0  \\
        -s_{\alpha_H} & c_{\alpha_H} & 0\\
        0 & 0 & 1 \\
    \end{array}\right)
    \left(\begin{array}{c}
         h^{\text{SM}}  \\
         \widetilde{H}  \\
         \widetilde{A}  \\
    \end{array}\right),
\end{equation}
where $c_{\alpha_c}=\cos\alpha_c, s_{\alpha_c}=\sin\alpha_c$, etc. So, the discovered 125 GeV Higgs $h$ may not behave exactly as SM predictions. For example, consider Higgs-Gauge boson interactions:
\begin{equation}
\label{gauge}
    \mathcal{L}_{\text{kin}}~\supset~g^2 W_{\mu}^+W^{-\mu} \Phi_i^{\dag}\Phi_i~\supset~ \frac{1}{2}g^2 W_{\mu}^+W^{-\mu} v h^{\text{SM}}~=~\frac{1}{2}c_{\alpha_H} c_{\alpha_A} g^2  W_{\mu}^+W^{-\mu}v h.
\end{equation}
So the $hW_{\mu}^+W^{-\mu}$ coupling strength $g_{hWW}=\frac{1}{2}c_{\alpha_H} c_{\alpha_A}g^2v$ might be smaller than $\frac{1}{2}g^2v$ of the SM. The ATLAS Run 2 data~\cite{ATLAS:2022vkf} tells $c_{\alpha_H} c_{\alpha_A}\gtrsim 0.99~ (0.96)$ at 68\% (95\%) confidence level. In the limit $\alpha_H, \alpha_A\ll 1$, one can expand Eq~\ref{HiggsMix} and get a more concise expression: 
\begin{equation}
\label{expand}
    \begin{aligned}
        &h^{\text{SM}}~=~h-|\epsilon|(c_{\widetilde{\alpha}_c}H+s_{\widetilde{\alpha}_c}A),\\
        &\widetilde{H}+i \widetilde{A} ~=~\epsilon h + e^{i \alpha_c} (H+i A). 
    \end{aligned}   
\end{equation}
where $\epsilon=\alpha_H+i \alpha_A,~\widetilde{\alpha}_c=\alpha_c-\arg\epsilon$. In analog to $c_{\alpha\beta}$ in 2HDM~\cite{Gunion:2002zf}, $\epsilon$ encodes the decoupling limit so that one can rewrite it as $\widehat{\lambda}v^2/m_H^2$, where $\widehat{\lambda}$ is a dimensionless coupling. When $m_H$ is large enough, $\epsilon$ goes to zero so the discovered 125 GeV Higgs becomes exactly SM-like. On the other hand, since we are working within a nondecoupled theory, $m_H$ is bounded to around 500 GeV. As a consistency requirement, $\widehat{\lambda}$ must therefore be small, while $m_H$ is tending toward its maximal possible value. The low-energy theory likely lies right at the edge where the perturbative expansion breaks down, similar to recent concerns about the UV regime~\cite{Jarkovska:2023zwv, Milagre:2024wcg}. Fortunately, these potentially nonperturbative effects are flavor conserving and therefore do not overwhelm our predictions regarding flavor structures.

\subsection{The Low Energy Theory}
\label{Prediction}
Eq~\ref{mixHeavy} and \ref{expand} allow us to rewrite Eq~\ref{YukawaPhi} in physical basis concisely (by replacing $\rho_i+i\eta_i\subset\Phi_i $ with $\frac{v_i}{v} h^{\text{SM}}+\frac{u_{i}}{u }(\widetilde{H}+i\widetilde{A})$ and solving $Y_{10},\widetilde{Y}_{10},Y_{126}$ from Eq~\ref{fermionsmass}). After dropping superheavy states and higher-order terms, we get the tree-level interacting Lagrangian for the light neutral scalars and SM fermions: 
\begin{equation}
\label{YukawaPhys}
    \begin{aligned}
        -\mathcal{L}_{\Phi\overline{F}F}~\supset&~(\frac{m_E}{v}+\epsilon Y_E^{\ell\ell'}) h\overline{\ell_L} \ell'_R  + 
        (\frac{m_D}{v}+\epsilon Y_D^{q q'}) \overline{d^{q}_{L}} d^{q'}_{R} h+(\frac{m_U}{v}+\epsilon Y_U^{q q'})  \overline{u_L^{q}} u_R^{q'} h \\
        ~&~+\mathcal{Y}_E^{\ell\ell'} (H+i A)\overline{\ell_L} \ell'_R 
          +\mathcal{Y}_D^{q q'} (H+i A)\overline{d^{q}_{L}} d^{q'}_{R}  
          +\mathcal{Y}_U^{q q'} (H+i A)\overline{u_L^{q}} u_R^{q'} +\text{h.c.}
    \end{aligned}
\end{equation}
\begin{equation}
\label{SO10constrain}
    \begin{aligned}
       Y_E~&=~ {\cal C}_{EE}\frac{m_E}{v}~+~{\cal C}_{ED} V_E^*\frac{m_D}{v}V_E^{\dag}
       ~+~ {\cal C}_{EU}V_E^* V_{\text{CKM}}^T\frac{m_U}{v}V_{\text{CKM}}V_E^{\dag}, \\
       Y_D~&=~ {\cal C}_{DE}V_E^T\frac{m_E}{v}V_E~+~{\cal C}_{DD} \frac{m_D}{v}
       ~+~ {\cal C}_{DU} V_{\text{CKM}}^T\frac{m_U}{v}V_{\text{CKM}}, \\
       Y_U~&=~ {\cal C}_{UE}V_{\text{CKM}}^*V_E^T\frac{m_E}{v}V_EV_{\text{CKM}}^{\dag}~+~{\cal C}_{UD} V_{\text{CKM}}^T\frac{m_D}{v}V_{\text{CKM}}
       ~+~ {\cal C}_{UU}\frac{m_U}{v}, \\
    \end{aligned}
\end{equation}
where $\mathcal{Y}_F=e^{i\alpha_c}Y_F$, $\ell,\ell'=e,\mu,\tau$, $q,q'=d,s,b$ or $u,c,t$, running over the three families. Complex coefficients ${\cal C}_{FF'} (F,F'=E, D, U)$ read: 
\begin{equation}
    \label{coeff}
    \left(\begin{array}{ccc}
        {\cal C}_{EE}& {\cal C}_{ED}& {\cal C}_{EU} \\
        {\cal C}_{DE}& {\cal C}_{DD}&  {\cal C}_{DU}\\
        {\cal C}_{UE}  & {\cal C}_{UD} &  {\cal C}_{UU} \\
    \end{array}  \right)~=~\frac{v}{u}
    \left(\begin{array}{ccc}
        u_{1}& -u_{2}& -3 u_{3} \\
        u_{1}& -u_{2}&  u_{3} \\
        u_{2}^*  & u_{1}^* &  u_{4}^* \\
    \end{array}  \right)
    \left(\begin{array}{ccc}
        v_{10}^d& -v_{10}^{u*}& -3 v_{126}^d \\
        v_{10}^d& -v_{10}^{u*}&  v_{126}^d \\
        v_{10}^u& v_{10}^{d*}&  v_{126}^u \\
    \end{array}  \right)^{-1},     
\end{equation}
Direct calculation shows $\mathcal{C}_{EU}=\mathcal{C}_{DU}$, while we have also checked numerically that there are no other correlations among the elements in $\mathcal{C}_{FF'}$. One may assume the dimensionless parameters $\mathcal{C}_{FF'}$ are, in general, all around $O(1)$, but strictly speaking, they are not predicted. Large hierarchies of two or more orders of magnitude, in principle, are possible. For a better understanding, one can take the SUSY-like 2HDM potential~\cite{Branco:2011iw} as an analogy, where
\begin{equation}
    \mathcal{C}_{EE}~=~\mathcal{C}_{DD}~=~\tan\beta,\qquad \mathcal{C}_{UU}~=~\cot\beta,\qquad \text{others}~=~0. 
\end{equation}
There is only one theoretical constraint, that $\mathcal{C}_{FF'}$ should not be too large to ensure $\mathcal{Y}_F$ is within the limit of perturbative unitarity. The lower limit for $\mathcal{C}_{FF'}$ depends on how the fermions' mass relationship in Eq~\ref{fermionsmass} constrains $v_i$. It is likely that $\mathcal{C}_{FF'}$ cannot be all zero, but as discussed at the end of Sec.~\ref{Yukawa}, we cannot conclude a robust lower bound.

Our low energy theory has more freedom than the benchmark 2HDM without tree-level FCNC~\cite{Branco:2011iw}, while it is not the most general one, either. $SO(10)$ gives nontrivial constrains. Let us firstly try to intuitively understand it in the chiral limit. If $Y_{10}, \widetilde{Y}_{10}$ and $Y_{126}$ are all strictly zero, then the whole theory is invariant under the chiral transform $16_F\rightarrow 16_F e^{i\theta}$ and all fermions are massless. This is partly a good symmetry because, except for the top quark, all observed fermion masses are far smaller than the electroweak scale. Clearly, the large top quark mass $m_t$ explicitly breaks the chiral symmetry and requires large Yukawa couplings, but it is still reasonable to expect that many of the elements in $Y_E, Y_D, Y_U$ are small. In fact, if ignoring all fermion masses except for $m_t$, then we approximately have a tree-level result: 
\begin{equation}
\label{ChiralLimit}
\begin{aligned}
      Y_E^{\ell\ell'}~&\approx~\mathcal{C}_{EU}V_E^*V_{\text{CKM}}^{T}\text{Diag}(0,0,m_t) V_{\text{CKM}}V_E^{\dag}~\approx~\mathcal{C}_{EU}\frac{m_t}{v}(V_E^{\ell b}V_E^{\ell'b})^*,\\
      Y_D^{q q'}~&\approx~\mathcal{C}_{DU}V_{\text{CKM}}^{T}\text{Diag}(0,0,m_t) V_{\text{CKM}}~\approx~\mathcal{C}_{DU}\frac{m_t}{v}V_{\text{CKM}}^{t q}V_{\text{CKM}}^{t q'}, \\ 
       Y_U^{q q'}~&\approx~\mathcal{C}_{UU}\text{Diag}(0,0,m_t)~\approx~\mathcal{C}_{UU}\frac{m_t}{v}\delta^{q q'}.
\end{aligned}
\end{equation}
The small 2-3 and 1-3 mixing angles in the CKM matrix are assumed far smaller than the corresponding ones in $V_E$, to simplify $Y_E^{\ell\ell'}$. The flavor structure of $Y_E$ is next-to-minimal flavor Violation (NMFV)~\cite{Agashe:2005hk}, which can be checked by experiments. $Y_D$ then flavors minimal flavor violation (MFV)~\cite{DAmbrosio:2002vsn}, and $Y_U$ is approximately diagonal. It is clear that $\mathcal{C}_{EU}$ must be small, in order to avoid the tree-level FCNC in the leptonic sector. Remembering $\mathcal{C}_{EU}=\mathcal{C}_{DU}$, the quark sector is then automatically free from tree-level FCNC constrains. The theory can be made safe from FV constraints, mainly because most of the couplings are suppressed by $m_{\tau}/v, m_{b}/v$ and small mixing angles. The only specific requirement is a small $\mathcal{C}_{EU}$.

Eq~\ref{ChiralLimit} is only for illustration, and the hierarchies among $C_{FF'}$ elements might, in practice, be larger than $m_t/m_b$. However, there are still quantitative predictions on the nondiagonal flavor structure. Eq~\ref{fermionsmass} tells that $Y_{10}, \widetilde{Y}_{10}$ and $Y_{126}$ (and thus $Y_E, Y_D, Y_U$) are linear combinations of three symmetric fermion masses matrices $M_E, M_D,$ and $M_U$. Therefore, we have at tree level, 
\begin{equation}
\label{NonDiag}
\begin{aligned}
      Y_E^{\ell\ell'}~&\propto~(V_E^{\ell b}V_E^{\ell'b})^*+o(\lambda^2) (V_E^{\ell b}+V_E^{\ell'b})^*,  \quad \ell\neq\ell', \\
      Y_D^{q q'}~&\propto~V_E^{\tau q}V_E^{\tau q'}+\frac{\mathcal{C}_{DU}m_t}{\mathcal{C}_{DE} m_{\tau}} V_{\text{CKM}}^{t q}V_{\text{CKM}}^{t q'}, \quad  q\neq q', \\ 
       Y_U^{q q'}~&\propto~V_E^{\tau q}V_E^{\tau q'}+\frac{\mathcal{C}_{UD}m_b}{\mathcal{C}_{UE} m_{\tau}} V_{\text{CKM}}^{t q}V_{\text{CKM}}^{t q'}+o(\lambda), \quad q\neq q', 
\end{aligned}
\end{equation}
$Y_E^{\ell\ell'}, Y_D^{q q'}$ is valid up to $o(m_b/m_t)$ corrections. $\lambda\sim 0.2$ is the Wolfstein parameter, so $o(\lambda^2)$ encodes the small 2-3 and 1-3 mixing angles in the CKM matrix. It is clear that as long as $|V_{E}^{\ell b}|$ and $|V_{E}^{\ell'b}|$ are larger than $o(\lambda^2)$, the flavor structure for lepton flavor violation is indeed NMFV, shedding light on the unknown mixing matrix $V_E$. On the other hand, the flavor structure for down-type quark sector is not fully clear. One may expect that to be MFV, as of Eq~\ref{ChiralLimit}, but NMFV is also possible because $\mathcal{C}_{DU}$ should be suppressed to avoid FCNC. $Y_U^{q q'}$ may also receive $o(\lambda)$ corrections because $V_E^Tm_EV_E$ of Eq~\ref{SO10constrain} is sandwiched between $V_{\text{CKM}}^*$ and $V_{\text{CKM}}^{\dag}$.

\section{Phenomenology}
\label{Pheno}

\subsection{Is the theory safe?}
The second Higgs doublet has $\text{SU(2)}_L$ and $U(1)_{\text{Y}}$ charge by definition. Its interactions with weak gauge bosons change the electroweak precision data. The most sensitive one is $T$ parameter~\cite{Peskin:1990zt}; when the new scalars' masses are nearly degenerate (custodial limit), we have approximately~\cite{He:2001tp, Grimus:2007if}
\begin{equation}
\label{Tparameter}
     T~=~ \frac{(m_{H^+}^2-m_H^2)(m_{H^+}^2- m_A^2)}{ 48\pi s_w^2 m_W^2 m_H^2}~\approx~0.18\times \left(\frac{m_{H^+}-m_H}{100~ \text{GeV}}\right)\left(\frac{m_{H^+}-m_A}{100~ \text{GeV}}\right). 
\end{equation}
Fixing $U=0$, $ T$ should be smaller than 0.06 according to \cite{ParticleDataGroup:2024cfk}, excluding the CDF data~\cite{CDF:2022hxs}.\footnote{$1\sigma$ allowed CDF consistent range for $T$ is $ \{0.159, 0.210\}$~\cite{Babu:2022pdn}} The theory is safe although the second Higgs doublet is not heavy, because its contribution to the $T$ parameter is zero in the custodial limit. Consequently, the mass spectrum of $H^+, H, A$ is expected to be quasidegenerate.

Shall the second Higgs be directly observed at hardon colliders? It depends on the producing cross section and decay branching ratios. Assuming $o(1)$ $H\overline{t}t$ or $H\overline{b}b$ Yukawa couplings, the total cross section at the LHC is dominated by gluon-gluon fusion, and associated production with quarks, and may reach $1\sim10$ pb for $m_H\sim500$ GeV~\cite{Djouadi:2005gi, Djouadi:2005gj}. If the $H\overline{\tau}\tau$ Yukawa coupling is also  $o(1)$ (similar to minimal-super-symmetric SM with large $\tan\beta$), then the theory is excluded because the current collider constraints for $pp\rightarrow H  \rightarrow \overline{\tau}\tau$ are quite tight~\cite{CMS:2022goy}. $H\overline{\tau}\tau$ Yukawa couplings are therefore required to be smaller, leading to the final states dominated by $\overline{t}t$ and $\overline{b}b$ pairs. This scenario has large background at hadron colliders~\cite{Gaemers:1984sj,Dicus:1994bm}, so is not yet excluded~\cite{ATLAS:2024itc, ATLAS:2024vxm}. Additionally, $H$ may also be produced with the $W/Z$ vector bosons, and/or decay to $ZZ+WW$ final states. These modes are suppressed by $|\epsilon|^2$ ($c_{\alpha\beta}^2$, if using the notation of 2HDM benchmarks). A small $\epsilon$ is then also a required by the collider constraints, along with the Higgs precision measurements discussed in Sec.~\ref{Scalar}.

FCNC cannot be completely eliminated, but they do not necessarily exceed the experimental limits. Let us start with all $\mathcal{C}_{FF'}\sim o(1)$ as a benchmark. The neutral $B_s$ meson mixing induced by $Y_D$ is clearly too large. Consequently, $\mathcal{C}_{DU}$ must be suppressed by about
\begin{equation}
|\mathcal{C}_{DU}| ~\lesssim ~
\left( \frac{v}{m_t} \cdot \frac{1}{|V_{\text{CKM}}^{ts} V_{\text{CKM}}^{tb}|} \right) 
\times \frac{m_H / \sqrt{2}}{10^3~\text{TeV}} 
~\approx~ 0.013 \times \frac{m_H}{500~\text{GeV}},
\end{equation}
in the limit $m_H\approx m_A$. Here, $10^3$ TeV is the experimental limit for the cutoff of the dimension six operator $\mathcal{O}_4=\overline{b_L}q_R \overline{b_R}q_L$~\cite{Bona:2024bue}.  Fortunately, no further constraints for the other $\mathcal{C}_{FF'}$ are required. As a consequence of MFV, $B_d$ and $K$ mixing are less constrained than $B_s$ mixing. For the other $\Delta F=2$ processes, $D$ mixing is suppressed by $m_{\tau}/v$ and $m_{b}/v$, and the constrains from muonium-antimuonium oscillation is much weaker than neutral mesons. Given $|\mathcal{C}_{DU}|<10^{-2}, \mathcal{C}_{EU}=\mathcal{C}_{DU}$ and $\text{others}=o(1)$, all the $\Delta F=1$ processes for down-type quark and charged leptons, such as $B\rightarrow \mu^+\mu^-$ and $\mu\rightarrow e\gamma$, are also suppressed due to the lack of sizable chirality-flipping interactions. The top quark flavor-violating decay and anomalous production could be sizable, but the experimental sensitivities are significantly poorer.

\subsection{A window to check $SO(10)$}
\label{checkSO10}

\begin{figure}[t]
  \centering
  \includegraphics[width=0.50\textwidth]{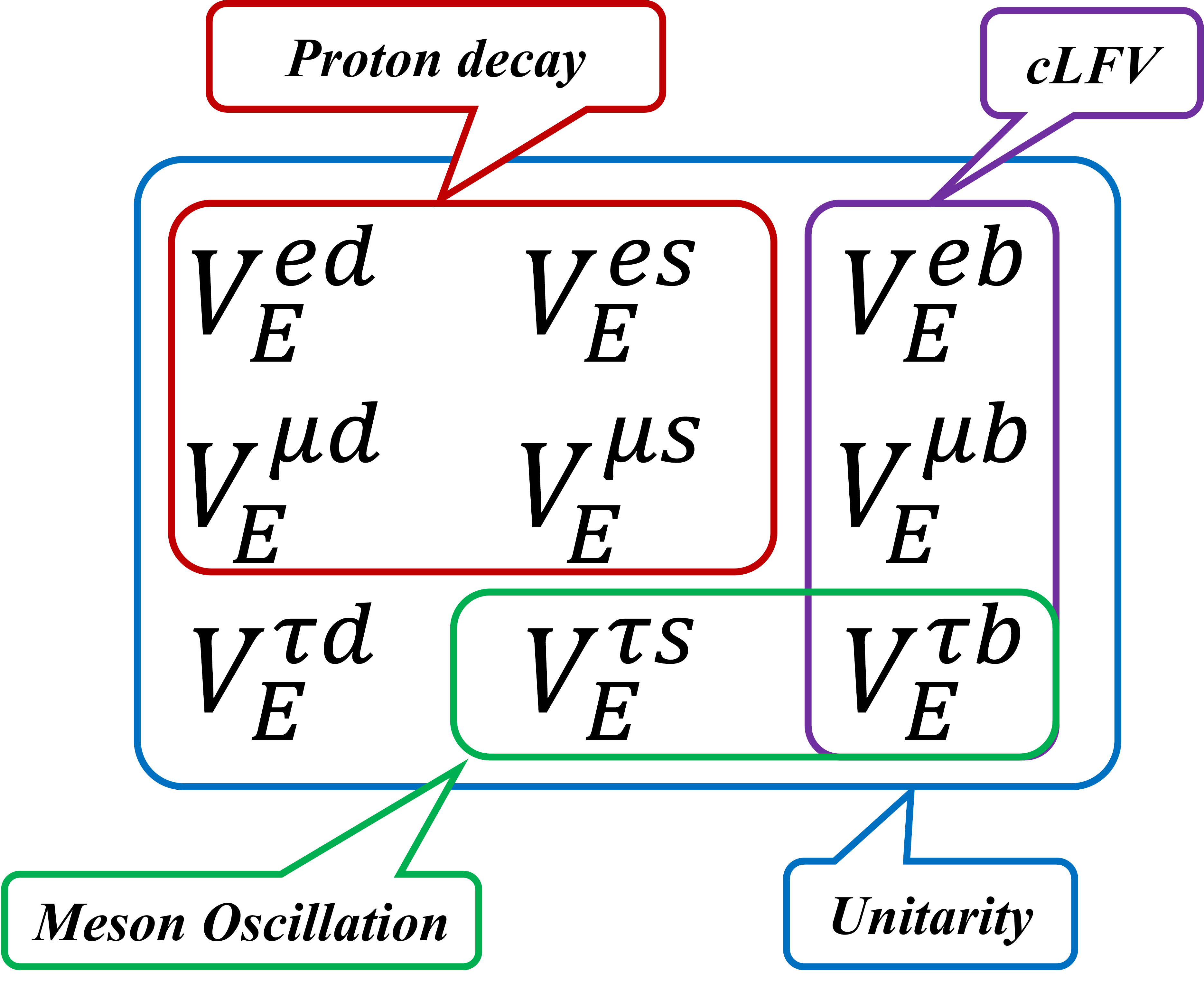}
    \caption{Proton decay, LFV, and neutral meson oscillation are linked through a $3\times3$ unitary matrix $V_E$. Together with the unitary conditions, the theory yields more observables than free parameters, leading to a nontrivial consistency check as we will show in Eq~\ref{expConnections}.}
   \label{VE}
\end{figure}

We now go to the predictions of $SO(10)$. The general idea is to demonstrate that there are more observables than free parameters. We focus on the absolute magnitudes of the unknown $3\times3$ mixing matrix $V_E$. Unitarity provides five independent constraints on the nine elements, $|V_E^{\ell q}|$, that the squared sum of any row or any column is one, minus the total. Physical predictions exist, as long as we can identify more than four direct experimental inputs. Various channels for lepton flavor violating (LFV) scalar decays, neutral meson oscillation, and proton decay are our candidates, as illustrated in Figure~\ref{VE}.

Eq~\ref{NonDiag} tells that LFV decay of $H$ and $A$ can decide the elements in the third column of $V_E$:
\begin{equation}
    \sigma(pp\rightarrow H,A)\times \text{Br}(H,A \rightarrow \ell\ell')~\propto~\left|Y_E^{\ell\ell'}\right|^2~\propto~|V_{E}^{\ell b}V_{E}^{\ell'b}|^2,
\end{equation}
valid as long as $|V_{E}^{\ell b}|$ and $|V_{E}^{\ell'b}|$ are larger than $o(\lambda^2)$. At colliders, we expect signals of heavy resonances decaying into $e\tau, \mu\tau, e\mu$ final states. A recent analysis is given in~\cite{CMS:2022fsw}. There will be two peaks in the $e\mu$ invariant mass spectrum, centered at $m_H$ and $m_A$. The prediction is about the relative total number of the excess events in each channel:  
\begin{equation}
\label{clfvelements}
    \frac{N_{e\mu}}{N_{\tau\mu}}~=~ \frac{|V_E^{eb}|^2}{|V_E^{\tau b}|^2}, \qquad 
    \frac{N_{e\mu}}{N_{e\tau}}~=~ \frac{|V_E^{\mu b}|^2}{|V_E^{\tau b}|^2}.
\end{equation}
Here, $N_{\ell\ell'}$ is the total number of the excess events, normalized by the detection efficiency. If measured in the future, then the unitarity of the third column of $V_E$, $|V_E^{eb}|^2+|V_E^{\mu b}|^2+|V_E^{\tau b}|^2=1$, would allow a direct extraction of the absolute values of all these three elements.

Elements in the third row of $V_E$ are related to quark FV. The down sector is better than the top quark one. It is partly because the top quark FV is much less precisely measured compared to neutral meson oscillations. The other reason is the nondiagonal elements of $Y_U$ receive $o(\lambda)$ correction from the CKM matrix, while those of $Y_D$ get only $o(\lambda^2)$ corrections. $h$ contribution is negligible in the small $\epsilon$ limit, and the dominate new physics contribution to the $\Delta F=2$ processes mainly comes from $H$ and $A$. The nondiagonal elements in $Y_D$ of Eq~\ref{YukawaPhys} gives an effective operator: 
\begin{equation}
\label{operator}
\begin{aligned}
    H_{\text{NP}}^q~&=~-\frac{1}{2m_H^2}\left(Y_D^{bq}\overline{b_L}q_R+Y_D^{qb*}\overline{b_R}q_L\right)^2
    -\frac{1}{2m_A^2}\left(iY_D^{bq}\overline{b_L}q_R-iY_D^{qb*}\overline{b_R}q_L\right)^2\\
    ~&\approx~-\frac{2|Y_D^{bq}|^2}{m_H^2}  \overline{b_L}q_R \overline{b_R}q_L, \quad  q=d,s. 
\end{aligned}
\end{equation}
The Wilson coefficients for $\overline{b_L}q_R \overline{b_L}q_R$ and $\overline{b_R}q_L \overline{b_R}q_L$ vanish in the limit $m_H=m_A$~\cite{Crivellin:2013wna}. This quasidegenerate limit is implied by the electroweak precision data, and can be cross checked by direct collider search. The operator $\overline{b_L}q_R \overline{b_R}q_L$ then dominantes, whose coefficient $Y_D^{bq}Y_D^{qb*}$ is nearly real because $Y_D$ is symmetric at tree-level. $H_{\text{NP}}^q$ is therefore approximately $CP$ conserving, and connected to experimental observables via
\begin{equation}
\label{hq}
    \begin{aligned}
        \frac{\langle B_q^0 |H_{\text{NP}}^q | \overline{B_q^0}\rangle}{\langle B_q^0|H_{\text{SM}}^q|\overline{B_q^0}\rangle }~=~ \textbf{h}_{\textbf{q}} e^{i\sigma_q}~=~\frac{M_{12}^{q\text{Exp}}}{M_{12}^{q\text{SM}}}-1, \quad q=d,s.
    \end{aligned}
\end{equation}
$2|M_{12}^{q\text{Exp}}|=\Delta m_q$ is the oscillation frequency. $\textbf{h}_{\textbf{q}}$ is the experimental input for our work, accounting for possible NP contributions. In pure SM, $\textbf{h}_{\textbf{d}}$ and $\textbf{h}_{\textbf{s}}$ are exactly 0 by definition. According to~\cite{Charles:2020dfl}, The current 95\% CL upper limits for $\textbf{h}_{\textbf{d}}$ and $\textbf{h}_{\textbf{s}}$ are, respectively, $0.26$ and $0.12$, 
and the future sensitivities at high-intensity experiments can reach about $0.03\sim0.04$. We summarize them in Table~\ref{clfv}.

\begin{table}[t]
    \caption{Current experimental bounds and future expected sensitivities on the FV processes relevant for our analysis. The constraints for $\textbf{h}_{\textbf{d}}$ and $\textbf{h}_{\textbf{s}}$, the parameters on how much $B_d$ and $B_s$ meson mixing derive from SM predictions, are shown with 95\% CL, using the results of CKMfit~\cite{Charles:2020dfl}. The future sensitivity comes from LHCb and Belle II, with 300/fb and 250/ab integrated luminosity respectively. A further possible improvement with $5\times10^{12}$ Z decays from a lepton collider like FCC-ee~\cite{FCC:2018evy} is included in the bracket. $(\Delta M_K)_{\text{NP}}$ is the maximally allowed NP contribution to the $K_L-K_s$ mass difference, where the SM long distance contribution is based on lattice QCD~\cite{Bai:2014cva, Wang:2022lfq}. We take the assumption of Snowmass 2021~\cite{Blum:2022wsz} that the future lattice errors can achieve 5\% level, for future sensitivity. All the observables in the Table are zero, if the SM is exact.}
    \vspace{10pt}    
\renewcommand{\arraystretch}{1.3}
    \centering
    \begin{tabular}{l l l  }  
    \hline
    Observable & Current limit & Future sensitivity \\
    \hline
    $\textbf{h}_{\textbf{d}}$ & 0.26~\cite{Charles:2020dfl} & 0.049~\cite{Charles:2020dfl} ~(0.038~\cite{FCC:2018evy}) \\
    $\textbf{h}_{\textbf{s}}$ & 0.12~\cite{Charles:2020dfl} & 0.044~\cite{Charles:2020dfl} ~(0.031~\cite{FCC:2018evy})\\
    $|(\Delta M_K)_{\text{NP}}|$ & $5.2\times 10^{-15}~\text{GeV}$~\cite{KTeV:2010sng, Wang:2022lfq} & $0.2\times 10^{-15}~\text{GeV}$~\cite{KTeV:2010sng,Blum:2022wsz}\\
    \hline
    \end{tabular}
    \label{clfv}
\end{table}

$H,A$ also contribute to $K$ meson mixing:
\begin{equation}
\label{MK}
    (\Delta M_K)_{\text{NP}}~=~\frac{1}{m_K} \langle K^0 |H_{\text{NP}}^K | \overline{K^0}\rangle~\approx~-\frac{2|Y_D^{sd}|^2}{m_K m_H^2}  \langle K^0 |\overline{d_L}s_R \overline{d_R}s_L| \overline{K^0}\rangle. 
\end{equation}
The $CP$ violating parameter $\epsilon_K$ receives zero contribution in the limit $m_H=m_A$. The hadronic factors, along with those for $B$ mesons, are shown in Appendix~\ref{Hadronic}. The SM prediction for $\Delta m_K$ is partly swamped by the long distance contributions, and the effect from the additional Higgs is not necessarily subdominant. The current measurement on $(\Delta M_K)$ is $3.484(6)\times10^{-15}$ GeV, and is not likely to improve much in future~\cite{KTeV:2010sng}. The latest lattice calculation shows $(\Delta M_K)_{\text{SM}}=5.8(0.6)_{\text{stat}}(2.3)_{\text{sys}}\times10^{-15}$ GeV~\cite{Wang:2022lfq, Bai:2014cva}, allowing sizable NP contribution with a sign opposite to the SM estimation. It is expected that reducing SM prediction uncertainty to the 5\% level is possible in the future, with adequate computer resources~\cite{Blum:2022wsz}.

\begin{figure}[!t]
  \centering
  \includegraphics[width=0.7\textwidth]{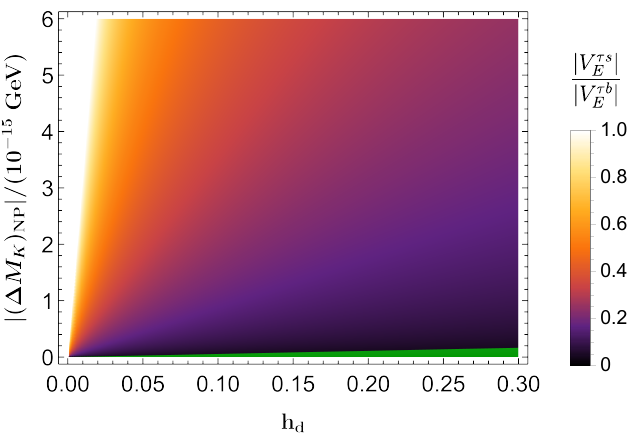}
    \caption{$|(\Delta M_K)_{\text{NP}}|$ versus $\textbf{h}_{\textbf{d}}$, where the dependence of $|V_E^{\tau s}|/|V_E^{\tau b}|$ is represented by the color scale on the right. The white color indicates $|V_E^{\tau s}|/|V_E^{\tau b}|\gtrsim1$, implying large mixing angles in $V_E$. The relationship of Eq~\ref{BNMFV} generally holds, except for the narrow green band of $|V_E^{\tau s}|/|V_E^{\tau b}|\lesssim\lambda^2\sim 0.04$. Parameter extraction becomes challenging when $\textbf{h}_{\textbf{d}}\rightarrow0$ and $|(\Delta M_K)_{\text{NP}}|\rightarrow 0$, since SM is revealed in this limit.}
   \label{Bosci}
\end{figure}

Different from the leptonic sector, the FV structure for down-type quarks can be either MFV or NMFV. To distinguish, we notice the MFV scenario implies
\begin{equation}
\label{BMFV}
    \begin{aligned}
        \frac{|(\Delta M_K)_{\text{NP}}|m_{K}}{\langle K^0 |\overline{d_L}s_R \overline{d_R}s_L| \overline{K^0}\rangle}
        ~\approx~\lambda^4\cdot\frac{2\textbf{h}_{\textbf{d}}|M_{12}^{d\text{SM}}|m_{B_d}}{\langle B_d^0|\overline{b_L}d_R \overline{b_R}d_L|\overline{B_s^0}\rangle}
        ~\approx~\lambda^6\cdot\frac{2\textbf{h}_{\textbf{s}}|M_{12}^{s\text{SM}}|m_{B_s}}{\langle B_s^0|\overline{b_L}s_R \overline{b_R}s_L|\overline{B_s^0}\rangle}. 
    \end{aligned}    
\end{equation}
If no hierarchies are found, the flavor structure must be NMFV, and the mixing angles in $V_E$ are typically larger than $\lambda^2$. The prediction is 
\begin{equation}
\label{BNMFV}
\begin{aligned}
    \frac{|(\Delta M_K)_{\text{NP}}|}{2\textbf{h}_{\textbf{d}}|M_{12}^{d\text{SM}}|\xi_{B}}~=~
     \frac{|V_E^{\tau s}|^2}{|V_E^{\tau b}|^2}, \quad \text{with}\quad~\xi_B~=~ \frac{m_{B}\langle K^0 |\overline{d_L}s_R \overline{d_R}s_L| \overline{K^0}\rangle}{m_{K}\langle B_d^0|\overline{b_L}d_R \overline{b_R}d_L|\overline{B_d^0}\rangle}.
\end{aligned}
\end{equation}
This is valid as long as the left-hand side is measured much larger than $o(\lambda^4)$. We illustrate the physically interesting region in Figure~\ref{Bosci}. $|M_{12}^{d\text{SM}}|$ is referred from Eq~\ref{hq}, by taking $M_{12}^{q\text{Exp}}= 0.506 ~\text{ps}^{-1}$~\cite{HFLAV:2022esi} and the central value $\textbf{h}_{\textbf{d}}=0.075,~ \sigma_q=-1.4 $~\cite{Charles:2020dfl}. It is clear that for any nonzero value of $|(\Delta M_K)_{\text{NP}}|$ that can be determined by near future lattice calculation, NMFV is the only possible flavor structure for $Y_D$. MFV requires very small $|(\Delta M_K)_{\text{NP}}|$, as shown in the narrow green band. If both $|(\Delta M_K)_{\text{NP}}|$ and $\textbf{h}_{\textbf{d}}$ are sizable, then one can extract the third row of $V_E$ using the unitarity condition of $|V_E^{\tau b}|^2+|V_E^{\tau s}|^2+|V_E^{\tau d}|^2=1$, assuming $|V_E^{\tau d}|$ is already extracted from the LFV decay of $H$ or $A$.

The remaining unconstrained elements lie within the $2\times2$ top-left submatrix of $V_E$. At leading order, these terms do not appear in Eq~\ref{NonDiag} due to the suppression of $m_{\mu}/m_{\tau}$ and $m_s/m_b$. Fortunately, the fermion mass hierarchies are not related to the proton decay amplitudes mediated by the gauge bosons. In addition, it has long been known~\cite{FileviezPerez:2004hn} that the proton decay flavor structure can be significantly simplified when the Yukawa couplings are symmetric, as in our scenario. Therefore, using the symbols defined in Eq~\ref{fermionMixing}, the general mixing matrices $V_1\sim V_4, V_{UD}, V_{EN}$ of~\cite{Nath:2006ut} reduce to
\begin{equation}
    V_1~=~V_4~=~\mathds{1}, \quad V_2~=~V_3^{\dag}~=~V_E, \quad V_{UD}~=~V_{\text{CKM}}, \quad V_{EN}~=~V_{\text{PMNS}},
\end{equation}
at GUT scale. The neutrino channels are more predictive, as all three generations of neutrinos are summed over and $V_E$ is eliminated~\cite{FileviezPerez:2004hn}. If the intermediate symmetry of $SO(10)$ breaking is left-right or quark-lepton, then the tree-level masses for superheavy gauge bosons are degenerate~\cite{Bertolini:2012im}. So, the width for $p\to \pi^+ \overline{\nu}$ reads: 
\begin{equation}
    \Gamma(p\xrightarrow{}\pi^+\overline{\nu})~=~   \frac{g_{\text{GUT}}^2m_p}{2\pi M^4_{\text{GUT}}}  A^2\langle\pi^+|(du)_Rd_L|p\rangle^2.
\end{equation}
Here, $g_{\text{GUT}}$ is the $SO(10)$ gauge coupling at $M_{\text{GUT}}$, and $A$ is the QCD renormalization factor. Related hadronic matrix elements are listed in Appendix~\ref{Hadronic}. Similarly, the decay $p\to K^+ \overline{\nu}$ also does not depend on the flavor mixings, and the ratio
\begin{equation}
\label{pratio1}
\begin{aligned}
      \frac{\Gamma(p\xrightarrow{}\pi^+\overline{\nu})}{\Gamma(p\xrightarrow[]{}K^+\overline{\nu})}~=~&
    \frac{4\left(1-m_K^2/m_p^2\right)^{-2}\langle\pi^+|(du)_Rd_L|p\rangle^2}{\langle K^+|(us)_Rd_L|p\rangle^2+\lambda^2\langle K^+|(ud)_Rs_L|p\rangle^2}
    ~\approx~ 81.2, 
\end{aligned}
\end{equation}
is fixed. This relation helps us distinguish whether proton decay processes are dominated by vector gauge bosons or scalar leptoquarks. Compared with gauge interactions, scalar interactions maximally violate flavor symmetry, so verifying Eq~\ref{pratio1} implies the scalar contributions are negligible.

\begin{figure}[t]
  \centering
  \includegraphics[width=0.95\textwidth]{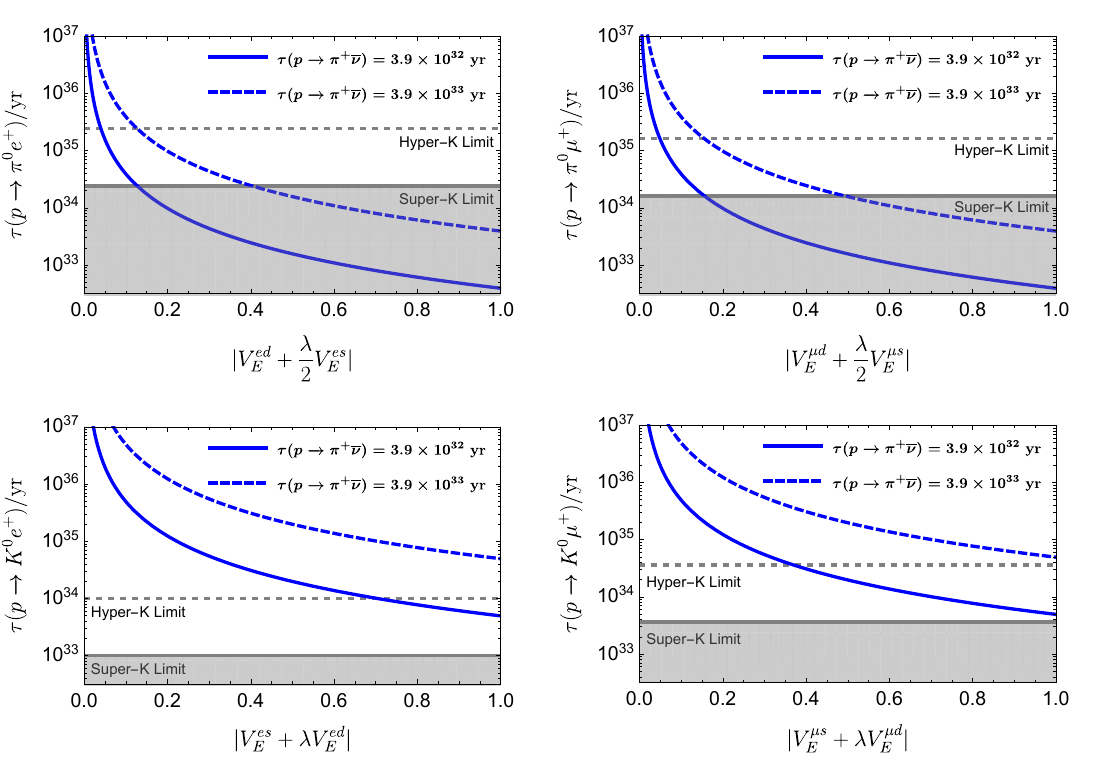}
    \caption{Lifetimes for various proton decay modes as a function of the magnitudes of elements in the $2\times2$ top-left submatrix of $|V_E|$. The blue dashed and solid lines represent the lifetimes in the benchmark of setting $\tau(p\xrightarrow{}\pi^+\overline{\nu})=3.9\times10^{32}$ years (current limit) or $\tau(p\xrightarrow{}\pi^+\overline{\nu})=3.9\times10^{33}$ years, respectively. The gray regions are excluded by Super-Kamiokande, as reported in Table~\ref{tab:pdecay}. Prospective Hyper-Kamiokande bounds are shown with the dashed gray lines, assuming an order of magnitude improvement in sensitivity. }
   \label{pdecayMixing}
\end{figure}

The branching ratios of proton decay to charged leptons are fixed up to $V_E$. Normalizing them with $\Gamma(p\xrightarrow{}\pi^+\overline{\nu})$ and $\Gamma(p\xrightarrow{}K^+\overline{\nu})$ dropping the $o(\lambda^2)$ terms in $V_{\text{CKM}}$, we have
\begin{equation}
\label{pdecayratio}
    \begin{aligned}
        \frac{\Gamma(p\xrightarrow{}\pi^0 e^+)}{\Gamma(p\xrightarrow{}\pi^+\overline{\nu})}~&=~|V_E^{ed}+\frac{\lambda}{2}V_E^{es}|^2, \qquad
        \frac{\Gamma(p\xrightarrow{}\pi^0 \mu^+)}{\Gamma(p\xrightarrow{}\pi^+\overline{\nu})}~=~|V_E^{\mu d}+\frac{\lambda}{2}V_E^{\mu s}|^2,\\ 
        \frac{\Gamma(p\xrightarrow{}K^0 e^+)}{\xi_K\Gamma(p\xrightarrow{}K^+\overline{\nu})}~&=~ ~|V_E^{es}+\lambda V_E^{ed}|^2, \qquad 
        \frac{\Gamma(p\xrightarrow{}K^0 \mu^+)}{\xi_K\Gamma(p\xrightarrow{}K^+\overline{\nu})}~=~
        ~|V_E^{\mu s}+\lambda V_E^{\mu d}|^2,\\ 
        \text{with}\quad~\xi_K~&=~\frac{2\langle K^0|(us)_Ru_L|p\rangle^2}{\langle K^+|(us)_Rd_L|p\rangle^2+\lambda^2\langle K^+|(ud)_Rs_L|p\rangle^2}~\approx~6.4.
    \end{aligned}
\end{equation}
The $SU(2)_L$ isospin symmetry for $u,d$ quarks is applied to simplify the hadronic elements related to neutral and charged pions. The latest measurement of different proton decay branching ratios are listed in Table~\ref{tab:pdecay}. Hyper-Kamiokande is expected to start operating as early as 2027 and improve the sensitivity by around one order of magnitude~\cite{Hyper-Kamiokande:2018ofw}. In Figure~\ref{pdecayMixing}, we show the partial lifetimes of various proton decay channels, setting the benchmarks of $\tau(p\xrightarrow{}\pi^+\overline{\nu})=3.9\times10^{32}$ years (current limit) or $\tau(p\xrightarrow{}\pi^+\overline{\nu})=3.9\times10^{33}$ years. The bounds from Super-Kamiokande and expected limits of Hyper-Kamiokande experiments are, respectively, indicated with solid and dashed gray lines. The diagrams suggest that, if $p\xrightarrow{}\pi^+\overline{\nu}$ is observed in near future, the lifetimes of proton decaying into charged leptons are generally within the limit of Hyper-Kamionkande, assuming elements in the $2\times2$ top-left submatrix of $|V_E|$ are sizable.

\begin{table}[t]
    \caption{90\% CL limits from proton decay searches on $\tau(p\to X)\equiv 1/\Gamma(p\to X)$. Future sensitivities are expected to be improved by about one order of magnitude at Hyper-Kamiokande~\cite{Hyper-Kamiokande:2018ofw}.  }
    \vspace{10pt}
\renewcommand{\arraystretch}{1.3}
    \centering
    \begin{tabular}{c c c c }
        \hline
        Decay mode & \(\ell=e^+\) & \(\ell=\mu^+\) & \(\ell=\bar{\nu}\) \\
        \hline
        \( p \to \pi \ell\) & \(> 2.4 \times 10^{34} \) yr~\cite{Super-Kamiokande:2020wjk} & \(> 1.6 \times 10^{34}\) yr~\cite{Super-Kamiokande:2020wjk} & \(> 3.9 \times 10^{32}\) yr~\cite{Super-Kamiokande:2013rwg} \\
        \( p \to K \ell \) & \(> 1.0 \times 10^{33}\) yr~\cite{Super-Kamiokande:2005lev} & \(> 3.6 \times 10^{33}\) yr~\cite{Super-Kamiokande:2022egr} & \(> 5.9 \times 10^{33}\) yr~\cite{Super-Kamiokande:2014otb} \\
        \hline
    \end{tabular}
    \label{tab:pdecay}
\end{table}

Eliminating the $o(\lambda)$ terms of Eq~\ref{pdecayratio}, we have
\begin{equation}
\label{pelements}
    \begin{aligned}
     \frac{2\Gamma(p\xrightarrow{}\pi^0 e^+)}{\Gamma(p\xrightarrow{}\pi^+\overline{\nu})} -\frac{\Gamma(p\xrightarrow{}K^0 e^+)}{\xi_K\Gamma(p\xrightarrow{}K^+\overline{\nu})}~&=~2|V_E^{ed}|^2- |V_E^{es}|^2,\\
     \frac{2\Gamma(p\xrightarrow{}\pi^0 \mu^+)}{\Gamma(p\xrightarrow{}\pi^+\overline{\nu})} -\frac{\Gamma(p\xrightarrow{}K^0 \mu^+)}{\xi_K\Gamma(p\xrightarrow{}K^+\overline{\nu})}~&=~2|V_E^{\mu d}|^2- |V_E^{\mu s}|^2.\\
    \end{aligned}
\end{equation}
This result is valid up to $o(\lambda^2)$. So similar to what we did before, assuming $|V_E^{eb}|,|V_E^{\mu b}|$ are extracted from cLFV processes, we can calculate all the elements in the $2\times 2$ top-left submatrix of $V_E$, using the unitarity relation of the first and second row, $|V_E^{ed}|^2+|V_E^{es}|^2+|V_E^{eb}|^2=1$ and $|V_E^{\mu d}|^2+|V_E^{\mu s}|^2+|V_E^{\mu b}|^2=1$.

In summary, we have identified five direct experimental observables, as shown on the left-hand side of Eq~\ref{clfvelements}, Eq~\ref{BNMFV}, and Eq~\ref{pelements}. They only depend on the absolute magnitudes of the elements of $V_E$, which can be parametrized by four free parameters. Therefore, if replacing $|V_E^{e b}|,|V_E^{\mu b}|, |V_E^{\tau b}|,$ and $|V_E^{\tau s}|$ in Eq~\ref{clfvelements} and Eq~\ref{BNMFV} by $|V_E^{ed}|,|V_E^{e s}|,|V_E^{\mu d}|,|V_E^{\mu s}|$, a concrete tree-level prediction emerges at leading order:
\begin{equation}
\label{expConnections}
\begin{aligned}
     &\frac{2\Gamma(p\xrightarrow{}\pi^0 \ell^+)}{\Gamma(p\xrightarrow{}\pi^+\overline{\nu})} -\frac{\Gamma(p\xrightarrow{}K^0 \ell^+)}{\xi_K\Gamma(p\xrightarrow{}K^+\overline{\nu})}\\
     ~=~&\left( \frac{3|(\Delta M_K)_{\text{NP}}|}{2\textbf{h}_{\textbf{d}}|M_{12}^{d\text{SM}}|\xi_B}  - \frac{N_{e\mu}}{N_{\tau\mu}}-\frac{N_{e\mu}}{N_{e\tau}}+1\right)\left( \frac{N_{e\mu}}{N_{\tau\mu}}+\frac{N_{e\mu}}{N_{e\tau}}+1\right)^{-1}, 
\end{aligned}
\end{equation}
where 
\begin{equation}
    \begin{aligned}
        \Gamma(p\xrightarrow{}\pi^0 \ell^+)~&=~\Gamma(p\xrightarrow{}\pi^0 e^+)+\Gamma(p\xrightarrow{}\pi^0 \mu^+),\\
        \Gamma(p\xrightarrow{}K^0 \ell^+)~&=~\Gamma(p\xrightarrow{}K^0 e^+)+\Gamma(p\xrightarrow{}K^0 \mu^+),\\
    \end{aligned}
\end{equation}
represents the total decay rate for both electron and muon final states. 
It is worth noting that this relation serves only as an illustration of an approximate constraint. Given various FCNC and proton decay modes are indeed measured in the future, we suggest fitting all the observables within the whole theory, for numerical robustness. The masses of the second generation fermions, and the mixing angles in the CKM matrix, should then also be included. To achieve better precision, it is also important to account for the RG effects. As shown in Appendix~\ref{rge}, the running toward, at least, the intermediate scale $M_I$ is well defined, since the light 2HDM parameters are directly measurable for near future experiments.
Verifying Eq~\ref{expConnections} in future experiments would indicate that all next-to-leading order corrections are indeed small, and thus provide a hint for minimal $SO(10)$. If minimal $SO(10)$ does not hold or the next-to-leading order corrections are overlooked, then the left- and right-hand sides of Eq~\ref{expConnections} can differ by orders of magnitude.

\section{Conclusion and discussion}
\label{conclu}

In this article, we have analyzed the predictions of the minimal realistic version of $SO(10)$ grand unified theory, that only one $CP$-even $45_H$, one $126_H$ and a complex $10_H$ are contained in the scalar sector. Additionally, we assumed $CP$ symmetry for all renormalized interactions, which in our scenario can only be spontaneously broken. An additional light Higgs doublet is required, as a consistency requirement. Different from general two-Higgs doublet models, the flavor structure of the predicted FCNC processes are connected to proton decay branching ratios, as a non-trivial constraint from matter unification.

The main uncertainty is the vacuum configuration of the Higgs doublets. They can not be clearly predicted from the scalar potential of $SO(10)$, and may receive large loop corrections due to the Higgs self-couplings. Perhaps, one can find some clues from fitting the charged fermion masses and mixing angles, but we do not get any reliable prediction. Therefore, we cannot estimate the absolute strengths of the FCNC signals. But fortunately, we can predict the relative strengths, provided various proton decay branching ratios are measured, since the flavor structure of minimal $SO(10)$ is quite constrained.

We acknowledge the widely known limit that the renormalizable $SO(10)$ GUT, if realistic, lacks predictive power. It does not allow direct calculation of the proton decay branching ratios. Meanwhile, there are no clear hints from the measured fermion masses and mixing angles. One possibility is that unification is simply not achieved, but leaving charge quantization as an unexplained coincidence. Alternatively, $SO(10)$ GUT might be nonminimal and all predictions are deeply hidden, so we can never prove it. However, we find minimal $SO(10)$ GUT may give a hint for electroweak scale physics, if GUT is slightly extended—by additionally assuming $CP$ is a fundamental symmetry, much like the common assumption of Lorentz invariance. 

The trouble—and possibly the reason this approach is overlooked—is the requirement of an additional fine-tuning. But, SM itself is also fine-tuned. The electroweak scale $v\ll M_{\text{GUT}}$ leads to $m_h\sim v\ll M_{\text{GUT}}$. The weak-scale S$CP$V requires two degenerate vacuum $v, v^*\ll M_{\text{GUT}}$, implying double fine-tuning and two light Higgs doublets. As long as one accepts non-SUSY GUT, there is no reason to reject the weak-scale S$CP$V. The fine-tuning puzzle does not necessarily change the low energy theory~\cite{Senjanovic:2020pnq}. Naturalness might be just hidden, for instance, via cosmological relaxation~\cite{Dvali:2019mhn}. If one considers the possibility that naturalness could be implicit, a class of well-motivated and potentially predictive models might remain unexplored. These theories, overlooked due to the naturalness criteria, or swamped in the supersymmetric models, could provide new insights. We thus advocate that theories without explicit naturalness also deserve thorough study in the future.

\section*{Acknowledgments}
The author thanks Goran Senjanović, Ulrich Nierste, and Robert Ziegler for useful discussions and comments on the manuscript. This research was partly supported the BMBF Grant No. 05H21VKKBA, \textit{Theoretische Studien für Belle II und LHCb.} X.G. also acknowledges the support by the Doctoral School ``Karlsruhe School of Elementary and Astroparticle Physics: Science and Technology.''

\section*{Data availability}
No data were created or analyzed in this study.

\newpage

\section*{Appendix}
\label{append}

\appendix

\section{Mixing matrix}
\label{mixing}
There are only three independent Yukawa couplings $Y_{10},\widetilde{Y}_{10}, Y_{126}$ in $SO(10)$ generic basis, so the mass matrices in Eq~\ref{fermionsmass} are connected via: 
\begin{equation}
\label{massconnection}
    M_{\nu_L}~=~ k_1 (M_E-M_D)+ k_2 M_U + k_3 M_U (M_E-M_D)^{-1} M_U, 
\end{equation}
where
\begin{equation}
  \begin{aligned}      
    k_1 ~=~ \frac{8(v_{126}^u)^2}{v_{126}^d\langle\Delta_R^0\rangle}- \frac{\langle\Delta_L\rangle}{8 v_{126}^d}, \quad
    k_2 ~=~ \frac{16v_{126}^u}{\langle\Delta_R^0\rangle}, \quad
    k_3 ~=~ \frac{8 v_{126}^d}{\langle\Delta_R^0\rangle}.
\end{aligned}  
\end{equation}
If rewriting Eq~\ref{massconnection} with physical masses and mixing matrices, it reads: 
\begin{equation}
\label{massconnectionP}
    V_E^T ( V_{\text{PMNS}}^* m_{\nu_L}^{\text{diag}} V_{\text{PMNS}}^{\dagger}- k_1 m_E^{\text{diag}})V_E ~=~
    -k_1 m_D^{\text{diag}} + k_2 (V_{\text{CKM}}^T m_U^{\text{diag}} V_{\text{CKM}})+k_3 M_T, 
\end{equation}
where
\begin{equation}
    M_T~=~V_{\text{CKM}}^T m_U^{\text{diag}} V_{\text{CKM}} (V_E^T m_E^{\text{diag}} V_E - m_D^{\text{diag}})^{-1}V_{\text{CKM}}^T m_U^{\text{diag}} V_{\text{CKM}}.
\end{equation}
The complexity of $M_T$ is the main barrier to solve $V_E$ from Eq~\ref{massconnectionP}. However, some rough approximation can shed light on its magnitudes. Neglecting $m_u, m_d, m_s$, and all mixings angles in the CKM matrix, we get:
\begin{equation}
\begin{aligned}
    M_T~&\approx~ \frac{m_{\tau}}{\det{M_T}}\times
    \left(
        \begin{array}{c c c}
           m_u^2 m_{b}\mathcal{D}_{11} & m_u m_c m_{b}\mathcal{D}_{12} & m_u m_t m_{\mu}\mathcal{D}_{13}\\
           m_um_c m_{b}\mathcal{D}_{21} & m_c^2 m_{b}\mathcal{D}_{22} & m_c m_t m_{\mu}\mathcal{D}_{23} \\
           m_u m_t m_{\mu}\mathcal{D}_{31} & m_c m_t m_{\mu}\mathcal{D}_{32} & m_t^2 m_{\mu}\mathcal{D}_{33}\\
        \end{array}
    \right),\\
        \mathcal{D}_{11}~&=~-(V_E^{\tau s})^2+ o(\frac{m_{\mu}}{m_b}), \qquad
        \mathcal{D}_{22}~=~-(V_E^{\tau d})^2+ o(\frac{m_{\mu}}{m_b}), \\
        \mathcal{D}_{33}~&=~(V_E^{\mu s}V_E^{\tau d}-V_E^{\mu d}V_E^{\tau s})^2+o(\frac{m_{s}}{m_{\mu}}), \qquad
        \mathcal{D}_{12}~=~\mathcal{D}_{21}~=~V_E^{\tau d}V_E^{\tau s}+ o(\frac{m_{\mu}}{m_b}), \\
        \mathcal{D}_{13}~&=~\mathcal{D}_{31}~=~(V_E^{\mu s}V_E^{\tau d}-V_E^{\mu d}V_E^{\tau s})(V_E^{\mu s}V_E^{\tau b}-V_E^{\mu b}V_E^{\tau s})+o(\frac{m_{s}}{m_{\mu}}), \\
        \mathcal{D}_{23}~&=~\mathcal{D}_{32}~=~(V_E^{\mu s}V_E^{\tau d}-V_E^{\mu d}V_E^{\tau s})(V_E^{\mu b}V_E^{\tau d}-V_E^{\mu d}V_E^{\tau b})+o(\frac{m_{s}}{m_{\mu}}).
    \end{aligned}
\end{equation}
Eq~\ref{massconnection} is valued at $M_{\text{GUT}}$, so we take the hierarchy $m_{\mu}\approx 5 m_{s}$ at $M_{\text{GUT}}$. This simplifies the leading order expression of $M_T$. Another important GUT scale relation is $m_t m_{\mu}\gg m_bm_{c}$. These relations give a clear hierarchy structure among the elements of $M_T$, assuming no elements of $\mathcal{D}_{ij}$ are accidentally small.

$M_T$ can be diagonalized with $M_T=V_T^*m_T^{\text{diag}}V_T^{\dag}$. This leads to $V_T^{ij}\sim 1$ when $i=j$ and $V_T^{ij}\ll 1$ when $i\neq j$. Therefore, the right-hand side of Eq~\ref{massconnectionP} can be diagonalized with a $3\times3$ unitary matrix closed to identity, assuming no accidental cancellations. The same flavor structure must also hold for the left-hand side, so the pattern of $V_E$ is clear: 
\begin{itemize}
    \item $k_1m_{\tau}\gg m_{\nu_L}\sim 0.1$ eV, $V_E\sim \mathds{1}$.
    \item $k_1m_{\tau}\lesssim m_{\nu_L}\sim 0.1$ eV, $V_E\sim V_{\text{PMNS}}$.
\end{itemize}
These two scenarios can be understood intuitively. Taking the limit that neutrinos are massless ($M_{\nu_L}=0$), Eq~\ref{massconnection} becomes a constraint for $M_E, M_D, M_U$. Ignoring the nondiagonal elements of the CKM matrix, both $M_D, M_U$ can be taken as diagonal, and then $M_E$ is automatically diagonal in this limit. Taking into account the neutrino masses, the nondiagonal elements of $V_E$ can be large, potentially as large as ones in the maximal flavor mixing matrix $V_{\text{PMNS}}$.

It is important to note that the discussion above is not robust. A careful reader may have already noticed that even with moderate cancellation, the predictions could change significantly.
In fact, such cancellation is somehow motivated, since $\langle\Delta_R^0\rangle$ is expected to lie around $10^{11}$ GeV (or $10^{9}$ GeV) for successful gauge coupling unification without fine-tuned light scalars. But the correct light neutrino mass implies a larger value, $\langle\Delta_R^0\rangle\sim10^{13}$ GeV. If $v_{126}^u$ and $v_{126}^d$ are not simultaneously suppressed, then one need some accidental cancellation among the $(3,3)$ elements of the three matrices on the right-hand side of Eq~\ref{massconnection}.
We discuss the potential structure of $V_E$ here, mainly to give readers a feeling about how much information is available from the fermion masses and mixing angles.

\section{Renormalization group effects}
\label{rge}

We illustrate how the RG effects change the symmetric structure of the Yukawa matrices here. The running equations between the GUT scale $M_{\text{GUT}}$ and the intermediate scale $M_I$ remain indeterminate because the physics around $M_I$ is not directly accessible by experiments. Although the minimal scalar spectrum around $M_I$ is well defined, the related Yukawa couplings remain unknown. Numerical fits may provide some insights, but these couplings cannot be faithfully derived from theory or directly measured by the experiments. Only the running equations from $M_I$ down to the electroweak scale (SM with light 2HDM) are well defined, assuming no other particles much lighter than $M_I$. 

The explicit running equations for general 2HDM can be found in~\cite{Branco:2011iw}. While our low energy theory is somehow specified, significant simplifications are possible. Firstly, terms that only change the Yukawa couplings by an overall factor can be dropped at one loop. Moreover, within the the parameter space of interest, only the (3,3) elements of $Y_U$ and $M_U/v$ are large as of $o(1)$, while all other Yukawa couplings are suppressed. Furthermore, the 2-3, 1-3 mixing angles of $V_{\text{CKM}}$ and the mass ratios $(m_c/m_t), (m_s/m_b), (m_{\mu}/m_{\tau})$ can all be neglected as an approximation. As a result at one loop, we conclude that at the electroweak scale $M_{\text{EW}}$\footnote{As an illustration, we take the masses of the electroweak gauge bosons, top quark, and all physical states of light 2HDM (supposed to be smaller than 500 GeV) equal to $M_{\text{EW}}$.}
\begin{equation}
\label{RGeq1}
\resizebox{0.93\textwidth}{!}{
$\begin{aligned}
    (Y_E^{\tau \ell}-Y_E^{\ell\tau})\big|_{M_{\text{EW}}}~=&~ ( Y_E^{\tau \ell}-Y_E^{\ell\tau})\big|_{M_I}, 
        \qquad (Y_U^{t q'}-Y_U^{q' t})\big|_{M_{\text{EW}}}~=~ ( Y_U^{t q'}-Y_U^{q' t})\big|_{M_I},\\
        (Y_D^{b q}-Y_D^{q b})\big|_{M_{\text{EW}}}~=&~ ( Y_D^{bq}-Y_D^{qb})\big|_{M_I}+ Y_D^{bq}\times  \frac{y_t^2}{16\pi^2}\log\left(M_I/M_{\text{EW}} \right)\left(\frac{3}{2}|\mathcal{C}_{UU}|^2-\frac{1}{2}\right),\\
        (M_D^{\tau \ell}-M_D^{\ell \tau})\big|_{M_{\text{EW}}}~=&~ (M_D^{\tau \ell}-M_D^{\ell \tau})\big|_{M_I}, \qquad 
        (M_D^{tq'}-M_D^{q't})\big|_{M_{\text{EW}}}~=~ (M_D^{tq'}-M_D^{q't})\big|_{M_I},\\
        (M_D^{bq}-M_D^{qb})\big|_{M_{\text{EW}}}~=&~ (M_D^{bq}-M_D^{qb})\big|_{M_I}+(m_b V_E^{\tau b}V_E^{\tau q})\times \frac{y_t^2}{16\pi^2}\log\left(M_I/M_{\text{EW}}\right)\left(\mathcal{C}_{UU}\mathcal{C}_{DE}\frac{2m_{\tau}}{m_b}\right).  \\
\end{aligned}
$
}
\end{equation}
Here $\ell=e,\mu,~q=d,s,~q'=u,c$, and for simplicity we took NMFV, a necessary condition to extract $V_E^{\tau q}$,  as the flavor structure of $Y_D$. Notably, both $Y_D, M_D$ are always not exactly symmetric at the low scale, even if all the Yukawa couplings are accidentally symmetric at $M_I$.\footnote{Without specifying the physics at $M_I$, one cannot quantify the deviations of $Y_U, Y_E, M_U, M_E$. Those may be similar to $Y_D, M_D$, or fully negligible.} Taking $M_I\sim 10^{11}$ GeV and $y_t=\frac{\sqrt{2}m_t}{v}\approx 0.5$ at $M_I$, we estimate:
\begin{equation}    
\frac{y_t^2}{16\pi^2}\log\left(M_I/M_{\text{EW}}\right)~\approx~3\%.
\end{equation}
The unknown parameters $\mathcal{C}_{UU}$ and $\mathcal{C}_{DE}$ must be smaller than $o(1)$ due to the FCNC constraints. Consequently, we expect percent-level corrections to the Wilson coefficients of $\overline{b_L} q_R \overline{b_R} q_L$. Likewise, some mixing angles relevant for proton decay, zero at GUT scale, are shifted by a few percent of $V_E^{\tau b}V_E^{\tau q}$ at low energy. These corrections should be added to Eq~\ref{expConnections} once future experiments measure the parameters $\mathcal{C}_{UU}$ and $\mathcal{C}_{DE}$.

It is worth mentioning that the Yukawa textures at $M_I$, in general, still differ with those at the $M_{\text{GUT}}$. And more corrections such as QCD ($1+o(\alpha_s/4\pi)$) should be multiplied to $(Y_F-Y_F^T)\big|_{M_I}$ and $(M_F-M_F^T)\big|_{M_I}$ of Eq~\ref{RGeq1}, as two loop order corrections to the GUT scale result. However, we believe that the one-loop running effects from $M_I$ down to $M_{\text{EW}}$ are already a good illustration in practice, due to the logarithmic enhancement $\log\left(M_I/M_{\text{EW}}\right)$. Nevertheless, it seems quite unlikely for the full RG effects to be large. If the minimal $CP$ conserving $SO(10)$ is indeed the UV theory, then we expect the GUT scale predictions hold reasonably well at low energy.

\section{Hadronic elements}
\label{Hadronic}

The hadronic matrix elements related to proton decay are given in~\cite{Aoki:2017puj}:
\begin{equation}
\begin{aligned}
\langle K^0 | (us)_R u_L |p\rangle ~&=~ 0.103(3)(11)\,\text{GeV}^2, \quad  &
\langle K^0 | (us)_L u_L |p\rangle ~&=~ 0.057(2)(6)\,\text{GeV}^2, \\
\langle K^+ | (us)_R d_L |p\rangle  ~&=~ -0.049(2)(5)\,\text{GeV}^2,  &
\langle K^+ | (ud)_R s_L |p\rangle ~&=~ -0.134(4)(14)\,\text{GeV}^2,\\
\langle \pi^+ | (du)_R d_L |p\rangle ~&=~ -0.186(6)(18)\,\text{GeV}^2. \\
\end{aligned}
\end{equation}

For neutral meson oscillation, the corresponding hadronic matrix elements are commonly parametrized with bag parameters~\cite{Gabbiani:1996hi}. We translate the latest lattice results~\cite{Dowdall:2019bea, Boyle:2024gge} into the following explicit expressions:
\begin{equation}
\label{Bhadronic}
\begin{aligned}
    \langle K^0 |\overline{d_L}s_R \overline{d_L}s_R| \overline{K^0}\rangle~&=~-0.039~\text{GeV}^4,\quad  
    & \langle K^0 |\overline{d_L}s_R \overline{d_R}s_L| \overline{K^0}\rangle~&=~ 0.088~\text{GeV}^4, \\
    \langle B_d^0|\overline{b_L}d_R\overline{b_L}d_R| \overline{B_d^0}\rangle~&=~-0.52~ \text{GeV}^4, 
    &\langle B_d^0 |\overline{b_R}d_L\overline{b_L}d_R | \overline{B_d^0}\rangle~&=~0.96~ \text{GeV}^4, \\
    \langle B_s^0 |\overline{b_L}s_R\overline{b_L}s_R| \overline{B_s^0}\rangle~&=~-0.84~ \text{GeV}^4,
    & \langle B_s^0 |\overline{b_R}s_L\overline{b_L}s_R | \overline{B_s^0}\rangle~&=~1.40~ \text{GeV}^4. 
\end{aligned}
\end{equation}
The hadronic elements for $B$ mesons are at the physical point, and those for kaons are at $\mu=3$ GeV. The uncertainties are typically at a few percent level. We do not renormalize them to $m_H\sim 500$ GeV because only the ratios contribute to the physical prediction, which are independent of the renormalization scale.

\newpage

\bibliographystyle{JHEP}
\bibliography{SO10_CPV_ref.bib}

\end{document}